\documentclass[journal,comsoc]{IEEEtran}
\usepackage{amsmath,amsfonts}
\usepackage{algorithmic}
\usepackage{algorithm}
\usepackage{array}
\usepackage[caption=false,font=normalsize,labelfont=sf,textfont=sf]{subfig}
\usepackage{textcomp}
\usepackage{stfloats}
\usepackage{url}
\usepackage{verbatim}
\usepackage{graphicx}
\usepackage{cite}

\usepackage{tikz}
\usetikzlibrary{arrows,graphs,shapes}
\usepackage{comment}
\usepackage{amsthm}
\usepackage{mathrsfs}

\def\listsize{L}
\def\AuxiliaryProb{Q}
\def\visitedleaves{\mathcal{V}}
\def\unvisitedleaves{\mathcal{W}}
\def\fbinary{\mathbb{F}_2}

\def\frozenset{\mathcal{F}}
\def\nonfrozenset{\mathcal{A}}

\def\llr{\mathscr{L}}
\def\UERthres{\varepsilon}

\def\infosize{M}
\def\outersize{K}
\def\innersize{N}
\def\logtwoinnersize{n}
\def\kernelmatrix{F^{\otimes \logtwoinnersize}}

\def\activeset{\mathcal{X}}

\newtheorem{thm}{Theorem}

\newtheorem{lemma}[thm]{Lemma}

\begin{document}
\title{Improving the decoding performance of CA-polar codes\\
{}
\thanks{This work was supported by the Defense Advanced Research Projects Agency (DARPA) under Grant HR00112120008, and by the National Science Foundation under Grant ECCS-2433994 and ECCS-2433996}
}

\author{Jiewei~Feng, Peihong Yuan, Ken R. Duffy and Muriel M\'edard
\thanks{J. Feng and K. R. Duffy, Northeastern University, (e-mails: \{feng.ji,k.duffy\}@northeastern.edu).}
\thanks{P. Yuan and M. M{\'e}dard, Massachusetts Institute of Technology, (e-mails: \{phyuan, medard\}@mit.edu).}
}

\maketitle

\begin{abstract}
We investigate the use of modern code-agnostic decoders to convert CA-SCL from an incomplete decoder to a complete one.
When CA-SCL fails to identify a codeword that passes the CRC check, we apply a code-agnostic decoder that identifies a codeword that satisfies the CRC. We establish that this approach gives gains of up to $0.2$ dB in block error rate for CA-polar codes from the 5G New Radio standard. If, instead, the message had been encoded in a systematic CA-polar code, the gain improves to more than $1.5$ dB. Leveraging recent developments in blockwise soft output, we additionally establish that it is possible to control the undetected error rate even when using the CRC for error correction. 
\end{abstract}


\section{Introduction}
\IEEEPARstart{P}{olar} codes were introduced by Ar{\i}kan in 2008 and are capacity achieving for binary-input discrete memoryless channel (B-DMC) \cite{Arikan09polar}.
When concatenated with a cyclic redundancy check (CRC) as an outer code, the resulting CRC-Aided polar (CA-polar) codes were adopted in the 5G New Radio (NR) standard for all control channel communications \cite{Bioglio23_5G}. CRC-Aided Successive Cancellation List (CA-SCL) decoding, e.g.~\cite{Niu12CASCL,tal15list}, is a prominent decoder for CA-polar codes. It operates by first providing a list of candidates using SCL decoding followed by performing CRC checks for each candidate. If no candidate passes the CRC check, a decoding failure is declared. A complete decoder always provides a codeword while an incomplete one does not \cite{blahut2003algebraic}. As a result, CA-SCL is an incomplete decoder.

While the CRC in CA-polar codes is used for error detection, it can be more extensively leveraged for error correction using recent advances. Guessing Random Additive Noise Decoding (GRAND)~\cite{Duffy19,solomon20, duffy2022_ordered} and Guessing Codeword Decoding (GCD)~\cite{ma2024guessing,zheng2024universal} are code-agnostic decoders that can efficiently decode a broad class of codes, including CRC codes. These approaches have variants that achieve Maximum Likelihood (ML) decoding or closely approximate it. The algorithms can be highly parallelized, and many efficient circuits and chips have been published for GRAND variants, 
e.g. \cite{Mohsin2021VLSIORBGRAND,condo2022fixed,blanc2024GRANDABDecoder,Riaz25ORBGRAND,abbas2025improved,Kizilates25ORBRGANDAI}. In addition, they can be further accelerated using techniques from \cite{rowshan2022_constrained,rowshan2025segmented,Lukas2025Tree}. In this paper, we investigate the application of one of these recent code-agnostic decoders as an outer decoder that uses the CRC code for error correction when CA-SCL fails. This results in a complete decoder that we call Complete CA-SCL (CCA-SCL).

As standard CA-polar codes are non-systematic, the log-likelihood ratios (LLRs) used as input for the outer decoder require further calculation based on channel LLRs. This calculation introduces additional challenges and motivates a more tailored framework. Compared to standard polar codes, systematic polar codes provide an advantage in terms of bit error rate (BER) while guaranteeing the same block error rate (BLER) \cite{Arikan11SystematicPolar}. The decoding process of systematic polar codes remains the same, except that the information bits are directly available in the encoded codeword. A systematic polar code concatenated with a CRC code results in a systematic CA-polar code. Our results demonstrate that additional improvement can be achieved using the proposed decoding scheme if 
a systematic CA-polar code is used.

One motivation for adopting CA-polar codes in 5G NR was to support Ultra-Reliable Low-Latency Communication (URLLC). Recently, soft output (SO) variants of GRAND, GCD, and SCL have been introduced, which are known as SOGRAND \cite{galligan2023upgrade,yuan2025SOGRAND}, SO-GCD \cite{duffy2025SOGCD} and SO-SCL \cite{yuan2025SOSCL}, respectively. These variants can provide blockwise SO in the form of an accurate estimate of the likelihood that each proposed decoding in a list is correct. The key enhancement over earlier approaches, e.g. \cite{Forney68}, is the development of a per-decoding estimate of the likelihood that the transmitted codeword is not in the decoded list. Such enhancement removes the requirement of list decoding while enabling near-optimal SO compared to an exhaustively computed SO~\cite{Jiewei2025BS}. In addition to using the outer decoder, we employ these SO algorithms to provide reliable SO for undetected error rate (UER) control, which facilitates the use of the CRC for error correction as well as detection. On the other hand, accurate SO for CA-SCL can be adapted from SO-SCL by leveraging the CRC, which enables flexible UER control for the proposed framework. In particular, this adaptation also applies to other decoders for general concatenated codes.

The remainder of this paper is organized as follows. Section \ref{sec_pre} introduces the basic setups. Section \ref{sec_pipeline} describes the CCA-SCL decoding pipeline, with simulation results given in Section \ref{sec_simulation}. A conclusion is given in Section \ref{sec_conclusion}.

\section{Preliminaries}\label{sec_pre}
Consider a binary message denoted as $\Phi^\infosize=(\Phi_1,\cdots,\Phi_\infosize)\in\fbinary^\infosize$ where $\infosize$ is the number of information bits. The encoding process of a CA-polar code can be described as follows. Let $G_{\text{O}}$ denote the $\infosize$ by $\outersize$ generator matrix for the systematic CRC outer code. The message $\Phi^\infosize$ is first encoded using this outer code through $\Phi^{\outersize}=\Phi^\infosize G_{\text{O}}$. $H_{\text{O}}$ denotes the parity check matrix of $G_{\text{O}}$. Let $G_{\text{I}}$ denote the $\outersize$ by $\innersize$ generator matrix of the polar code. The sequence $\Phi^\outersize$ is then encoded using this polar code as the inner code, through $X^{\innersize}=(X_1,\cdots,X_\innersize)=\Phi^{\outersize}G_{\text{I}}=\Phi^\infosize G_{\text{O}}G_{\text{I}}$. 

The polar code specified by $G_{\text{I}}$ can be equivalently described by the following: Let $\frozenset\subset\{1,\cdots,\innersize\}$ denote the locations of the frozen bits in a sequence with length $\innersize=2^n$ for some $n$ and $|\frozenset|=\innersize-\outersize$. We define $\nonfrozenset=\{1,\cdots,\innersize\}\setminus \frozenset$ where $\setminus$ is the set minus operation. Denote $G_{[\outersize,\innersize]}$ to be the $\outersize$ by $\innersize$ matrix such that the columns in the set $\frozenset$ form a $\outersize$ by $\innersize-\outersize$ zero matrix and the rest of the columns form a $\outersize$ by $\outersize$ identity matrix. The sequence $\Phi^\outersize$ is first expanded by $U^\innersize=\Phi^\outersize G_{[\outersize,\innersize]}$, where the subsequence $U_{\nonfrozenset}$ is given by $\Phi^\outersize$ and the subsequence $U_{\frozenset}$ is a zero sequence. Let $\kernelmatrix$ denote the $n$-th Kronecker power of $F$ and $F=\left[\begin{smallmatrix}1&0\\1&1\end{smallmatrix}\right]$. Then the expanded sequence is encoded by $\kernelmatrix$, resulting in $\Phi^\outersize G_{\text{I}}=\Phi^\outersize G_{[\outersize,\innersize]}\kernelmatrix$. Note that $\kernelmatrix$ is its own inverse.

After encoding, $X^\innersize$ is transmitted via $\innersize$ independent uses of a B-DMC, and the receiver observes channel output $Y^{\innersize}\in\mathbb{R}^\innersize$. Given knowledge of the channel, the conditional pdf of $Y$ given $X$, $f_{Y|X}$, can be calculated. The LLR of the $i$-th received bit is denoted as $\llr_{\text{I},i}=\log[{f_{Y|X}(Y_i|X_i=1)}/{f_{Y|X}(Y_i|X_i=0)}].$ We denote $\llr_{\text{I}}=(\llr_{\text{I},1},\cdots,\llr_{\text{I},\innersize})$, where $\text{I}$ is used to indicate that the LLR is related to the $\innersize$ bits resulting from encoding through the inner code.

An SCL decoder \cite{tal15list} decodes and provides a list of $L$ candidates $\hat{U}^{\innersize,1},\hat{U}^{\innersize,2},\cdots,\hat{U}^{\innersize,\listsize}$ for $U^\innersize$. Let $\visitedleaves$ denote the set of candidates. A LLR-based SCL decoder provides a path metric (PM) \cite{balatsoukas2015llr} for each of the candidates to indicate which candidate is more reliable. For each candidate $\hat{U}^{\innersize,i}$, SO-SCL \cite{yuan2025SOSCL} provides a corresponding SO, $S_i\in[0,1]$, which serves as a probability estimate for whether the candidate is correct, i.e., whether $\hat{U}^{\innersize,i}={U}^{\innersize}$. Note that the ordering of the PM for each candidate is the same as the ordering of the SO for each candidate. Let $\textbf{0}$ denote the zero vector. Each candidate in $\visitedleaves$ is examined by the CRC check. Denote $\visitedleaves'=\{\hat{U}^{\innersize,i}|\hat{U}^{\innersize,i}\in\visitedleaves,H_{\text{O}}(\hat{U}^{\innersize,i}_{\nonfrozenset})^T=\textbf{0}\}$ to be the subset of $\visitedleaves$ such that $\visitedleaves'$ contains all the candidates that pass the CRC check.  Let $S^*=\max\{S_i|\hat{U}^{\innersize,i}\in\visitedleaves'\}$ denote the maximum SO within the remaining candidates and the corresponding candidate $\hat{U}^{\innersize,*}\in\{\hat{U}^{\innersize,i}|S_i=S^*\}\subseteq\visitedleaves'$ will be the decoded message. If $\visitedleaves'=\emptyset$, a decoding failure is declared.

\section{Proposed pipeline for CCA-SCL}\label{sec_pipeline}
\usetikzlibrary{arrows}
\tikzstyle{arrow}=[
    thick,
    ->,
    >=stealth
    ]
    
\begin{figure}
\centering
\begin{tikzpicture}
    \node (0) at (-2,0) {};
    \node (1) [rectangle,draw,fill=green!20!] at (0,0) {SO-SCL};
    \node (2) [diamond,draw,aspect=2,align=center,fill=blue!20!] at (4,0) {CRC\\check};
    \node (31) [rectangle,draw,align=center] at (-0.5,-2) {Output $\hat{U}^{\innersize,*},S^*$};
    \node (32) [rectangle,draw,align=center,fill=yellow!20!] at (4.5,-1.7) {Calculate LLRs for $\Phi^{\outersize}$ \\ using $\llr_{\text{I}}$};
    \node (41) [diamond,draw,aspect=3,fill=blue!20!] at (-0.5,-3.3) {$S^*>1-\UERthres$?};
    \node(42) [rectangle,draw,fill=green!20!] at (4.5,-2.75) {SO-GRAND/SO-GCD};
    \node (51) [rectangle,draw,align=center] at (4.5,-3.6) {Output $\hat{\Phi}^{\outersize,*},S^*$};
    \node (61) [rectangle,draw] at(2,-4.1) {Terminate};

    \draw [arrow,blue] (0)-- node[anchor=south] {$\llr_{\text{I}}$} (1);
    \draw [arrow,blue] (1) -- node[anchor=south] {$\visitedleaves$}node[anchor=north,black]{$S_1,\cdots,S_\listsize$} (2);
    \draw [arrow,blue] (2.south) -- (4,-1)  -| node[anchor=north]{$\quad\quad\quad\quad\visitedleaves'\neq\emptyset\quad$} (31.north);
    \draw [arrow] (2.south) --(4,-1)-| node[anchor=west]{$\visitedleaves'=\emptyset$} (32);
    \draw [dashed,arrow] (31) -- (41);
    \draw [dashed,->,thick,orange] (41.east)--node[anchor=east,black]{No}(32.west);
    \draw [arrow] (32)--(42);
    \draw [arrow] (42)--(51);
    \draw [dashed,arrow] (51)--(41.east);
    \draw [dashed,arrow] (41.south)--(-0.5,-4.1)--node[anchor=south]{Yes}(61.west);
\end{tikzpicture}
\caption{Pipeline of the proposed CCA-SCL decoder.}
\label{fig_Pipeline1}
\end{figure}

Here we investigate a decoding scheme based on CA-SCL to improve BLER with accurately controllable UER, which is illustrated in Fig. \ref{fig_Pipeline1}. SO-SCL first decodes as explained in Section \ref{sec_pre}. When $\visitedleaves'\neq\emptyset$, the most reliable candidate is selected. These operations are similar to a traditional CA-SCL decoder except that SCL is replaced by SO-SCL in order to provide accurate SO for UER control. In Fig. \ref{fig_Pipeline1}, the processes for standard CA-SCL decoding are indicated in blue. 

When $\visitedleaves'=\emptyset$, CA-SCL declares a failure. Instead, we apply a decoder for the outer code by using the LLRs corresponding to $\Phi^{\outersize}=U_{\nonfrozenset}$ and $H_{\text{O}}$, which will be described in detail in Section \ref{sec_converting}. The decoder for the outer code is called the \textit{outer decoder}, and possible choices of the outer decoder include SO-GCD \cite{duffy2025SOGCD} and SOGRAND\cite{yuan2025SOGRAND}. The procedure of applying the outer decoder is demonstrated on the right side of Fig. \ref{fig_Pipeline1}. The outer decoder provides a decoded candidate $\hat{\Phi}^{\outersize,*}$ with corresponding SO $S^*$. The proposed pipeline, regardless of whether $\visitedleaves'$ is empty, always provides a valid codeword $\hat{U}^{\innersize,*}$ or $\hat{\Phi}^{\outersize,*}$ with its SO $S^*$, which enables additional operations that are highlighted by dashed arrows in Fig. \ref{fig_Pipeline1}. 

One new possibility is to perform a threshold test of the SO for UER control, as highlighted in black dashed lines. By adjusting the threshold $\epsilon$, it is possible to control the tradeoff between BLER and UER. For accurate SO, the tradeoff is optimal in the sense that there is no other decision rule that can have lower BLER and UER simultaneously as established in \cite{Forney68}. A second possibility is to activate the outer decoder whenever the decoded codeword from CA-SCL fails the threshold test, which is highlighted in the orange line.

\subsection{Decoding using outer decoder}\label{sec_converting}
When CA-SCL fails to decode, we apply an outer decoder to estimate $\Phi^\outersize$ with the following inputs: LLRs corresponding to $\Phi^\outersize=U_{\nonfrozenset}$ called \textit{outer LLR} and the CRC parity check matrix $H_{\text{O}}$. However, the outer LLR cannot be directly observed from $\llr_{\text{I}}$ since the polar code is not systematic, leading to the following calculation. Conditioned on the channel LLR, the likelihood that $X_i=1$, denoted as $B_{\text{I},i}$, can be calculated as
\begin{align}
   B_{\text{I},i}=P(X_i=1|\llr_{\text{I}})=\frac{1}{1+e^{-\llr_{\text{I},i}}}.\label{eq_Bi}
\end{align} 

Let $\pi(i)$ denote the position in $U^\innersize=\Phi^{\outersize}G_{[\outersize,\innersize]}$ that stores the value of $\Phi_i$. Let $\kernelmatrix_{i,j}$ denote the entry of $\kernelmatrix$ at the $i$-th row and the $j$-th column. Since $\Phi^{\outersize}G_{[\outersize,\innersize]}\kernelmatrix=\Phi^{\outersize}G_{\text{I}}=X^\innersize$, $\Phi^{\outersize}G_{[\outersize,\innersize]}=U^\innersize=X^\innersize\kernelmatrix$. Hence the likelihood that $\Phi_i=1$ conditioned on the channel LLR can be calculated using Lemma $1$ of \cite{Gallager62}, i.e., 
\begin{align}
    B_{\text{O},i}=P(\Phi_i=1|\llr_{\text{I}})=\frac{1}{2}-\frac{1}{2}\prod_{j=1}^{\innersize}\left(1-2B_{\text{I},j}\kernelmatrix_{j,\pi(i)}\right),\label{eq_converting}
\end{align}
which can be converted to outer LLR $\llr_{\text{O},i}$ for the $\outersize$ bits in $\Phi^\outersize$ analogously to eq. \eqref{eq_Bi}. This calculation leads to a new set of LLRs, which are characterized in Lemmas \ref{lem_cov} and \ref{lem_lower_rel}.

\begin{lemma}\label{lem_cov} Denote $\activeset_i=\{s|\kernelmatrix_{s,\pi(i)}=1\}$ so that $\Phi_i=\bigoplus_{s\in\activeset_i}X_s$ where $\bigoplus$ denotes the summation in $\fbinary$. $\activeset_i$ represents the set of bits in $X^\innersize$ that can be used to recover $\Phi_i$. Let $\activeset_{i,j}=\activeset_i\cap \activeset_j$. Define $p_{i\setminus j}:=\frac{1}{2}-\frac{1}{2}\prod_{s\in\activeset_{i}\setminus \activeset_{i,j}}(1-2B_{\text{I},s})$ and $p_{i\setminus j}:=0$ if $\activeset_i\subseteq \activeset_j$. Define $p_{i,j}:=\frac{1}{2}-\frac{1}{2}\prod_{s\in\activeset_{i,j}}(1-2B_{\text{I},s})$ and $p_{i,j}:=0$ if $\activeset_{i,j}=\emptyset$. Then the covariance $\text{Cov}(\Phi_i,\Phi_j|\llr_{\text{I}})=0$ if and only if one of the following holds: 1) $p_{i,j}\in\{0,1\}$; or 2) $p_{i\setminus j}=\frac{1}{2}$ or $p_{j\setminus i}=\frac{1}{2}$.
\end{lemma}

\begin{proof}
    We will omit the notation for conditioning on $\llr_{\text{I}}$ throughout the proof as all calculations will be done with the conditioning. Given a pair $i$ and $j$, notice that $\Phi_i=\left(\bigoplus_{s\in \activeset_i\setminus \activeset_{i,j}} X_s\right)\oplus\left( \bigoplus_{s\in  \activeset_{i,j}} X_s\right)$ and similarly for $\Phi_j$. Let $\bigoplus_{s\in \activeset} X_s=0$ if $\activeset=\emptyset$. The same approach as in eq. \eqref{eq_converting} leads to $P\left( \bigoplus_{s\in  \activeset_{i,j}} X_s=1\right)=p_{i,j}$, $P\left(\bigoplus_{s\in \activeset_i\setminus \activeset_{i,j}} X_s=1\right)=p_{i\setminus j}$, and $P\left(\bigoplus_{s\in \activeset_j\setminus \activeset_{i,j}} X_s=1\right)=p_{j\setminus i}$.
    The covariance can then be calculated directly, yielding $\text{Cov}(\Phi_i,\Phi_j)=p_{i,j}(1-p_{i,j})(1-2p_{i\setminus j}-2p_{j\setminus i}+4p_{i\setminus j}p_{j\setminus i})$.
\end{proof}

Condition 1) is satisfied when the demodulated bits with indices in $\activeset_{i,j}$ have infinite reliability or $\Phi_i,\Phi_j$ are affected by disjoint sets of $X^\innersize$ so that $\activeset_{i,j}=\emptyset$. However, in polar coding, $\activeset_{i,j}\neq \emptyset$ by the construction of $\kernelmatrix$. Therefore, condition 1) happens with probability zero. Condition 2) is satisfied when all the bits with indices in $\activeset_{i\setminus j}\cup\activeset_{j\setminus i}$ have reliabilities equal to zero, which has probability zero. Therefore, after converting the channel LLRs to the outer LLRs, the converted LLRs are correlated, resulting in degradation if the outer decoder assumes $\Phi_i,\Phi_j$ are independent for all $i,j$. In addition, the outer LLR has a smaller magnitude, as stated in Lemma \ref{lem_lower_rel}.

\begin{lemma}\label{lem_lower_rel}
Let $|\llr_{\text{I},j}|$ denote the reliability of the hard decision for $X_j$ based on $\llr_{\text{I},j}$. Similarly, let $|\llr_{\text{O},i}|$ denote the reliability of the hard decision for $\Phi_i$ based on $\llr_{\text{O},i}$. Then 
$|\llr_{\text{O},i}|\leq|\llr_{\text{I},j}|$ for all $i,j$ such that $j\in\activeset_i$.
\end{lemma}

\begin{proof}
    By eq. \eqref{eq_Bi},
        $\left|1/2-B_{\text{I},i}\right|=({e^{|\llr_{\text{I},i}|}-1})/({2e^{|\llr_{\text{I},i}|}+2})$,
which implies that $|1/2-B_{\text{I},i}|$ is a monotone increasing function of the reliability for the corresponding bit. A parallel statement holds for $|1/2-B_{\text{O},i}|$. Hence it suffices to show $|1/2-B_{\text{O},i}|\leq |1/2-B_{\text{I},j}|$ for all $i,j$ such that $j\in\activeset_i$. Note that
$\left|1/2-B_{\text{O},i}\right|
    =2^{|\activeset_i|-1}\prod_{j\in\activeset_i}\left|1/2-B_{\text{I},j}\right|
        \leq  \left|1/2-B_{\text{I},j}\right|$
    for all $j$ such that $j\in\activeset_i$, which completes the proof.
\end{proof}

Lemmas~\ref{lem_cov} and \ref{lem_lower_rel} indicate a potential challenge when using the outer LLRs for decoding. Here, we demonstrate that systematic polar codes~\cite{Arikan11SystematicPolar} can be used instead for further improvement. The encoding process of a systematic polar code can be done by methods described in~\cite{Sarkis2016SystematicPolar,Sarkis2014FastPolarDecoders}, which enables the outer LLRs to be directly observed from the channel LLRs. First, the message $\Phi^\outersize$ is encoded for the first time to get $U^\innersize=\Phi^\outersize G_{\text{I}}$. Then only the part $U_{\nonfrozenset}$ is encoded for the second time by $X^\innersize=U_{\nonfrozenset}G_{\text{I}}$. The resulting $X_{\nonfrozenset}$ then stores the message $\Phi^\outersize$ with $X_{\frozenset}$ representing the redundant bits. An SCL decoder can decode with the same procedure as decoding non-systematic polar codes except that the information bits are now stored in $X_{\nonfrozenset}$ instead of $U_{\nonfrozenset}$. Applying systematic polar code in CA-polar, the CRC-encoded message is then stored in $X_{\nonfrozenset}=\Phi^\outersize$ instead, whose corresponding LLRs can be directly observed from $\{\llr_{\text{I},i}|i\in\nonfrozenset\}$, avoiding the degradation caused by conversion. Simulation results in Section \ref{sec_simulation} demonstrate the improvement achieved using systematic CA-polar codes.

\subsection{Managing UER with SO-SCL adapted to CA-polar}\label{SOCASCL}
When a decoded message is provided with a blockwise SO $S^*$, it can be used for UER control. A desired upper bound $\epsilon$ for UER can be achieved by rejecting any decoding with $1-S^*>\epsilon$. This is because a blockwise SO represents the probability estimate that the decoding is correct. In CCA-SCL, the outer decoder is used to provide blockwise SO for error detection through the threshold test. Although the CRC code is used for error correction in the proposed method, we demonstrate that error detection capability is maintained for the decoding produced by SCL. In this section, we will adapt the SO-SCL calculation to CA-SCL so that all decoded codewords produced in the pipeline are paired with a blockwise SO, which can be used to control UER.

We first give an overview of SO-SCL \cite{yuan2025SOSCL} for polar-like codes. We denote $U^{i}$ to be the subsequence of $U^{\innersize}$ such that $U^{i}=(U_1,\cdots,U_i)$. Assume the observed channel output is $Y^\innersize=y^\innersize$. Let $Q_{U^\innersize|Y^\innersize}$ denote an auxiliary conditional probability mass function (PMF) as described in ~\cite{yuan2025SOSCL}. The SC decoding performs a greedy search and identifies a sequence $\hat{u}^{\innersize}$ such that $\hat{u}^\innersize_\frozenset=\textbf{0}$. For SCL, up to $\listsize$ paths are explored in parallel, resulting in $\listsize$ candidates as output which are stored in $\visitedleaves$. The decoding process can be represented by a tree search with depth $\innersize$. Each path with length $\innersize$ in the tree represents a sequence $\hat{u}^\innersize\in\fbinary^\innersize$. SCL keeps a record of $\listsize$ paths that satisfy the frozen constraints, hence dividing the tree into three parts: a) \emph{Visited paths} denote the $\listsize$ paths provided by SCL, where all of them satisfy the frozen constraints; b) \emph{Unvisited valid paths} are the paths that satisfy the frozen constraints but have not been fully explored due to memory and complexity issues; c) \emph{Invalid paths} are the paths that do not meet the frozen constraints, where a path $\hat{u}^\innersize\in\fbinary^\innersize$ is invalid if $\hat{u}^\innersize_\frozenset\neq\textbf{0}.$

In this paper, the term \textit{valid} refers to whether the path satisfies the frozen constraints without considering the CRC. Recall that $\visitedleaves$ is the set of $\listsize$ visited paths. We denote $\unvisitedleaves$ to be the set of all unvisited valid paths. The optimal blockwise SO for polar-like codes with the knowledge of all possible codewords is given by $\Gamma\left(y^\innersize, \hat{u}^\innersize\right)
=$
\begin{align}
\frac{Q_{U^\innersize|Y^\innersize}\left(\hat{u}^\innersize\left|y^\innersize\right.\right)}{\sum\limits_{u^\innersize\in \visitedleaves} Q_{U^\innersize|Y^\innersize}\left(u^\innersize\left|y^\innersize\right.\right)+\sum\limits_{u^\innersize\in \unvisitedleaves} Q_{U^\innersize|Y^\innersize}\left(u^\innersize\left|y^\innersize\right.\right) }.\label{eq_SO_SCL_theory}
\end{align}

The numerator in (\ref{eq_SO_SCL_theory}) represents the likelihood of a decoded sequence $\hat{u}^\innersize$ while the denominator represents the sum of likelihoods of all possible decoding paths satisfying the frozen constraints. The denominator is called the codebook probability. Forney's approximation \cite{Forney68} to equation (\ref{eq_SO_SCL_theory}) is given by 
\begin{align}
    \frac{\AuxiliaryProb_{U^{\innersize}|Y^\innersize}(\hat{u}^\innersize|y^\innersize)}{\sum\limits_{u^\innersize\in\visitedleaves}\AuxiliaryProb_{U^\innersize|Y^\innersize}(u^\innersize|y^\innersize)}\label{eq_SO_Forney},
\end{align}
where the second summation in the denominator of (\ref{eq_SO_SCL_theory}) is omitted, i.e., the SO is calculated conditioned on the transmitted codeword being in the list of visited paths. Note that the assumption requires that the list size is at least two.

In contrast, \cite{yuan2025SOSCL} considers the omitted term in (\ref{eq_SO_Forney}), which represents the likelihood of unvisited valid paths. The omitted term $\sum_{u^\innersize\in \unvisitedleaves} Q_{U^\innersize|Y^\innersize}\left(u^\innersize\left|y^\innersize\right.\right) $ is approximated by a quantity $\AuxiliaryProb^*_{\unvisitedleaves}$ which can be readily calculated through the decoding process (see \cite{yuan2025SOSCL} for details). This leads to the approximation for (\ref{eq_SO_SCL_theory}) as $\Gamma\left(y^\innersize, \hat{u}^\innersize\right)$
\begin{align}
    &\approx\Gamma^*\left(y^\innersize, \hat{u}^\innersize\right)
=\frac{\AuxiliaryProb_{U^{\innersize}|Y^\innersize}(\hat{u}^\innersize|y^\innersize)}{\sum\limits_{u^\innersize\in\visitedleaves}\AuxiliaryProb_{U^\innersize|Y^\innersize}(u^\innersize|y^\innersize)+\AuxiliaryProb^*_{\unvisitedleaves}}.\label{eq_SOSCL}
\end{align}

It is verified in \cite{yuan2025SOSCL} that the blockwise SO using equation (\ref{eq_SOSCL}) is accurate and outperforms Forney's approximation for polar-like codes. Note that each $\hat{U}^\innersize$ corresponds to a unique sequence of $\hat{X}^\innersize$. This ensures that the same SO calculation can be applied in systematic polar codes.

For CA-polar codes, we can leverage the outer code to improve the accuracy of SO-SCL since not all the valid decoding paths in SCL can pass the CRC check. For any valid decoding path that violates the CRC check, its likelihood of being the transmitted codeword is zero. Recall that $\visitedleaves'\subseteq\visitedleaves$ denotes the set of visited paths that pass the CRC check. The summation $\sum_{u^\innersize\in\visitedleaves}$ in (\ref{eq_SOSCL}) is hence replaced by $\sum_{u^\innersize\in\visitedleaves'}$. 

Recall that the length of the CRC is $\outersize-\infosize$, and $\unvisitedleaves$ denotes the set of unvisited valid paths, which covers the $2^{\outersize}-L$ paths satisfying the frozen constraints. Note that $\AuxiliaryProb^*_{\unvisitedleaves}$ approximates the sum of likelihoods of the $2^{\outersize}-L$ unvisited valid paths regardless of the CRC check. Within the unvisited valid paths, around $[2^{\infosize}-L,2^{\infosize}]$ out of $2^{\outersize}-L$ of them are valid codewords that can pass the CRC check. We assume that the unvisited valid paths that satisfy the CRC check are uniformly distributed among all unvisited valid paths. This assumption may be affected when using static frozen bits, but is better satisfied by using random dynamic frozen bits \cite{yuan2025SOSCL}. Under this assumption, $2^{\infosize-\outersize}\AuxiliaryProb^*_{\unvisitedleaves}$ serves as an approximation for the sum of the likelihoods of the unvisited valid paths that can pass the CRC check. Combining the two modifications yields (\ref{eq_SOCASCL}), which provides the blockwise SO calculation for CA-polar codes using SO-SCL.

\begin{align}
        \Gamma^*\left(y^\innersize, \hat{u}^\innersize\right)=\frac{\AuxiliaryProb_{U^{\innersize}|Y^\innersize}(\hat{u}^\innersize|y^\innersize)}{\sum\limits_{u^\innersize\in\visitedleaves'}\AuxiliaryProb_{U^\innersize|Y^\innersize}(u^\innersize|y^\innersize)+2^{\infosize-\outersize}\AuxiliaryProb^*_{\unvisitedleaves}}.\label{eq_SOCASCL}
\end{align}

Note that equation (\ref{eq_SOCASCL}) only considers the length of the outer code $\outersize-\infosize$ and hence is compatible with any concatenated polar codes. More generally, for a concatenated code with outer code represented as a map $G_1:\fbinary^{\infosize}\rightarrow \fbinary^{\outersize}$ and an inner code represented as a map $G_2:\fbinary^{\outersize}\rightarrow \fbinary^{\innersize}$, where the outer code serves as an error detection code only and a decoder for the inner code is able to produce the blockwise SO using similar ideas from \cite{yuan2025SOGRAND,duffy2025SOGCD,yuan2025SOSCL}, a similar modification can be conducted to incorporate the outer code for SO calculation.

\section{Simulations}\label{sec_simulation}

\begin{figure}
    \centering
    \includegraphics[width=0.9\linewidth]{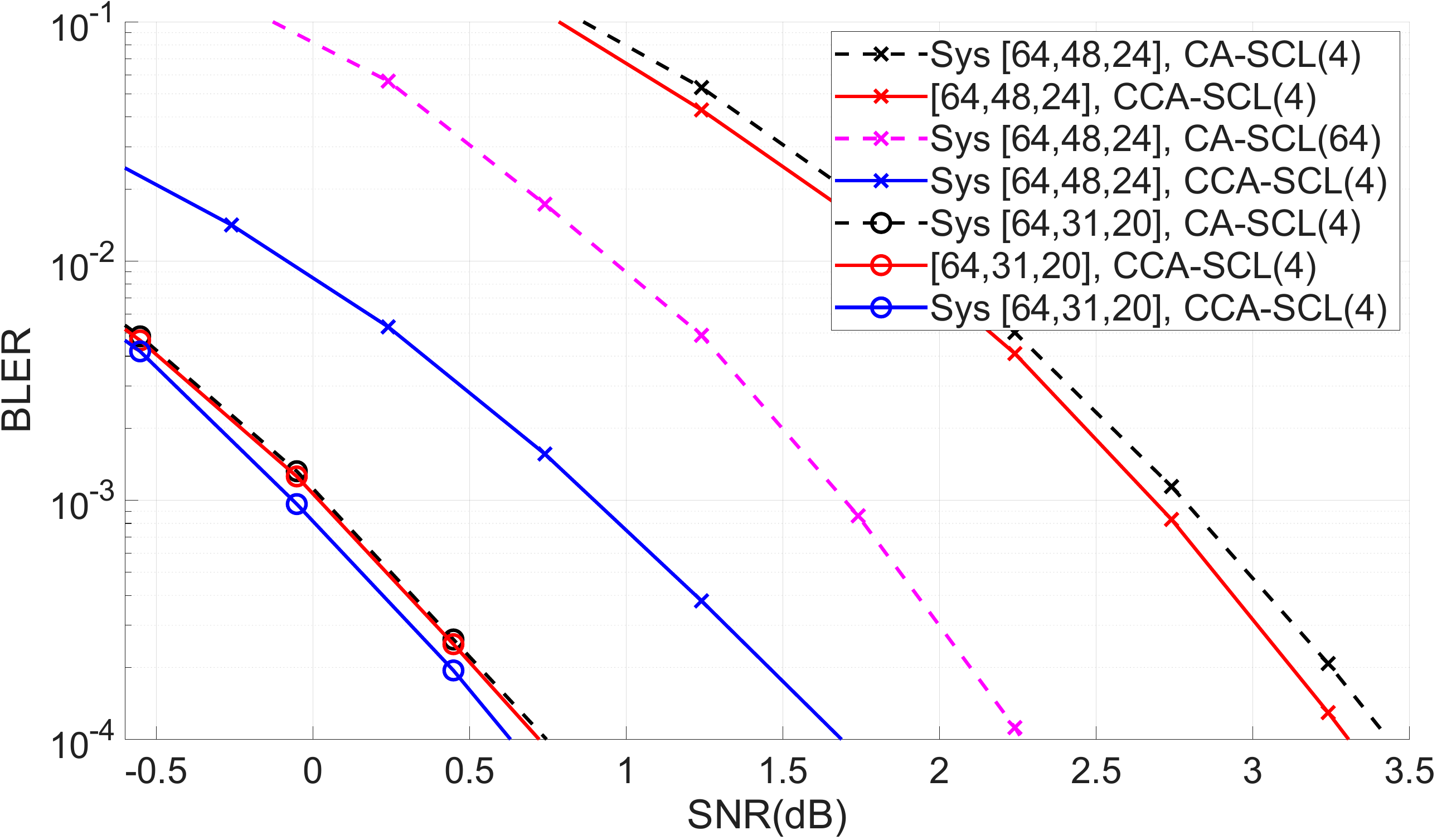}
    \caption{BLER performance of CA-SCL and Complete CA-SCL (CCA-SCL).}
    \label{fig_coummunicationlettters_combined}
\end{figure}

For simulations, we use BPSK and AWGN. The CRC parameters are chosen from the 5G NR standard. Fig. \ref{fig_coummunicationlettters_combined} demonstrates the BLER improvement of CCA-SCL without the threshold test. Since SO-GCD and SOGRAND have similar performance, we only present the results with SO-GCD as the outer decoder in this section. When decoded using CA-SCL, systematic CA-polar codes exhibit almost identical BLER to non-systematic CA-polar codes. Therefore, we only include results for systematic CA-polar codes when using CA-SCL as the decoder, which are represented by dashed lines. In the legend, the value in parentheses indicates the list size of the SCL decoder. Solid lines represent CCA-SCL where the outer decoder has list size one. Blue lines represent CCA-SCL decoding systematic CA-polar codes. Red lines represent CCA-SCL decoding non-systematic CA-polar codes. When decoding the $[\innersize,\outersize,\infosize]=[64,48,24]$ non-systematic CA-polar code, the outer decoder successfully identifies the correct codeword around $50\%$ of the time when CA-SCL fails. When the systematic CA-polar code is used instead, this success rate improves to approximately $99.7\%$. As depicted in Fig. \ref{fig_coummunicationlettters_combined}, CCA-SCL provides up to $0.2$ dB improvement for non-systematic codes and up to around $1.5$ dB gain for systematic codes.

SCL is known to have complexity $O(LN\log N)$. The likelihood that the outer decoder is activated is bounded above by CA-SCL's BLER, $p_e$. With $E[T]$ being the average outer decoder query count, which decreases as $SNR$ increases, the complexity of SO-GCD and SOGRAND have recently been characterized as $O(E[T](\outersize-\infosize)\infosize)$ and $O(E[T]\outersize+\outersize(\outersize-\infosize))$, respectively, \cite{qianfan2025BITS}. The complexity of CCA-SCL is, therefore, bounded above by $O(LN\log N+p_eE[T]\infosize(\outersize-\infosize)+p_e\outersize(\outersize-\infosize))$. 
The purple dashed line represents CA-SCL with a list size of $64$, serving as a benchmark for the $[64,48,24]$ systematic CA-polar code. It demonstrates that the gain achieved by the proposed method for systematic CA-polar codes can be difficult to recover even by substantially increasing the SCL list size. This is because the outer decoder leverages the CRC outer code for decoding. In addition, the outer decoder is only activated when SCL fails and operates on the shorter CRC code. This provides an additional advantage on top of the implementation efficiency of the outer decoders established in \cite{Mohsin2021VLSIORBGRAND,condo2022fixed,blanc2024GRANDABDecoder,Riaz25ORBGRAND,abbas2025improved,Kizilates25ORBRGANDAI}. 

\begin{figure}
    \centering
    \includegraphics[width=0.98\linewidth]{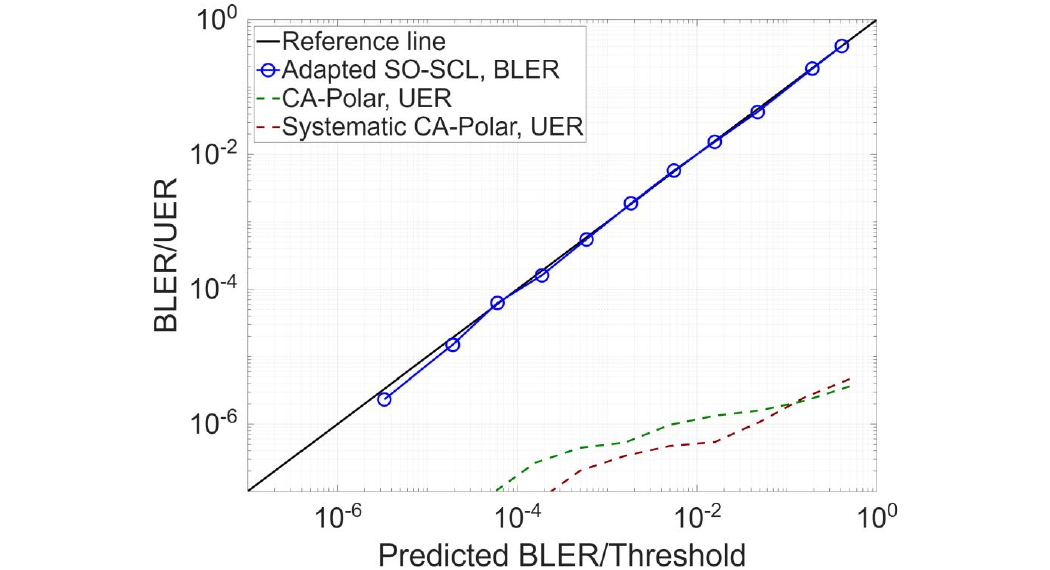}
    \caption{Verification of SO calibration for a $[64,43,32]$ CA-polar code.}
    \label{XYSO_64_32_11}
\end{figure}

As the outer decoder operates when SCL fails, it improves BLER at the cost of increasing UER. With the SO from the outer decoder and \eqref{eq_SOCASCL} for CA-SCL, the proposed method enables the UER to be maintained below a desired threshold. Compared to error detection through CRC outer codes, our proposed method is more flexible, as the targeted UER can be adjusted dynamically from the receiver side. In addition, \eqref{eq_SOCASCL} remains applicable even when the system is operated without the outer decoder. Fig. \ref{XYSO_64_32_11} provides verification of (\ref{eq_SOCASCL}) for CA-SCL and demonstrates the effectiveness of using SO for UER control in the pipeline. For verification of (\ref{eq_SOCASCL}), we use CA-SCL with a list size of eight to decode the $[64,43,32]$ CA-polar code at $E_b/N_0=2$ dB.  Simulations are grouped based on the predicted BLER $1-\Gamma^*(y^\innersize,\hat{u}^\innersize)$ within the intervals with endpoints $10^0,10^{-0.5},10^{-1},\cdots,10^{-5}$, where $\Gamma^*$ is calculated using (\ref{eq_SOCASCL}). For each group of simulations within an interval, the BLER is plotted against the average value of $1-\Gamma^*$. The black line represents the function $y=x$ as a reference. A plot similar to the black line indicates that the SO is accurate. The solid blue line represents (\ref{eq_SOCASCL}) and demonstrates its accuracy. Note that (\ref{eq_SO_Forney}) can also be adapted to CA-SCL by replacing $\visitedleaves$ with $\visitedleaves'$ in the summation of (\ref{eq_SO_Forney}). However, this modification cannot provide a meaningful SO when $|\visitedleaves'|=1$ because the decoded list size is one. In contrast, if (\ref{eq_SO_Forney}) is used directly, it provides an underestimated SO because it includes the sequence that violates the CRC check in the denominator. 

For verification of UER control, we employ the pipeline described in Fig. \ref{fig_Pipeline1}. Using the same code dimension and list size as above, we simulate at $E_b/N_0=5$ dB. UER versus the threshold $\epsilon$ is plotted. The dashed red line represents the simulation for non-systematic CA-polar while the dashed green line represents systematic CA-polar. Both lines are lower than the black line, which implies that CCA-SCL can maintain UER control within the targeted threshold.

\section{Conclusion}\label{sec_conclusion}
Employing an outer decoder when SCL fails, we introduce a complete decoding framework that can still control UER.
The CRC outer code is used for both error correction and detection, which enhances performance. Systematic polar encoding improves performance further by providing more reliable input to the outer decoder. These results suggest CCA-SCL can be deployed on existing systems for better performance.

\bibliographystyle{IEEEtran}
\bibliography{b.bib}

\end{document}